# Investigating Crowd Creativity in Online Music Communities

FABIO CALEFATO, University of Bari, Italy
GIUSEPPE IAFFALDANO, University of Bari, Italy
FILIPPO LANUBILE, University of Bari, Italy
FEDERICO MAIORANO, University of Bari, Italy

Crowd[1] creativity is typically associated with peer-production communities focusing on artistic products like animations, video games, and music, but less frequently to Open Source Software (OSS), despite the fact that also developers must be creative to come up with new solutions to their technical challenges.

In this paper, we conduct a study to further the understanding of which factors from prior work in both OSS and art communities are predictive of successful collaboration – defined as reuse of previous songs – in three different songwriting communities, namely Songtree, Splice, and ccMixter. The main findings from this study confirm that the success of collaborations is associated with high community status of recognizable authors and low degree of derivativity of songs.

## 1 INTRODUCTION

Crowd creativity is typically associated with peer-production communities in the arts domains, where members collaborate to create artistic products like animations, video games, and music [29]. At the same time, Open Source Software (OSS) is regarded as crowdsourcing applied to software development [21]. While defining OSS communities as creative may be arguable, we

Author's addresses: F. Calefato <fabio.calefato@uniba.it>, Dipartimento Jonico – Università degli Studi di Bari, Via Duomo, 259, 74123 - Taranto, Italy; G. Iaffaldano <giuseppe.iaffaldano@uniba.it>, Dipartimento di Informatica – Università degli Studi di Bari, via E. Orabona 4, 70125 - Bari, Italy; F. Lanubile <filippo.lanubile@uniba.it>, Dipartimento di Informatica – Università degli Studi di Bari, via E. Orabona 4, 70125 - Bari, Italy; F. Maiorano <f.maiorano2@studenti.uniba.it>, Dipartimento di Informatica – Università degli Studi di Bari, via E. Orabona 4, 70125 - Bari, Italy.

believe it is appropriate because: (i) writing software requires creativity when developers must come up with solutions to new technical challenges; (ii) there is an overlapping between OSS and arts communities. For example, communities like Newgrounds and Scratch revolve around the collaboration between artists, developers, musicians, voice actors, and writers in the creation of video games and animation.

In this work, we conduct a study to further our understanding of the factors influencing successful collaboration in online music communities. We note that the study is part of a broader research effort aimed to investigate collaboration, survivability, and retention in creative communities by applying the lessons learned from previous research on OSS communities.

There is a large body of prior work that has investigated success factors – i.e., the antecedents of successfully completed collaborations – in different types of online communities where, due to the differences in the artifact of interest, the perception of success varies. In this work, we focus on studying successful collaboration from the perspective of active users, that is, *creators* who collaborate actively in the peer production of shared artifacts (e.g., writing songs), as opposed to end-users, that is, *non-creators* who participate in the community exclusively by consuming them (e.g., playing songs). Albeit both creators and end-users are necessary for the survival of online creative communities, the presence of the latter is consequential to the availability of shared artifacts. Accordingly, hereinafter, a successful collaboration in online arts communities is intended in terms of artifact reuse, that is, the generation of derivative content through the reworking and recombination of existing contributions from other members [6, 13, 20].

In their study on the antecedents of reuse in the ccMixter music community, Cheliotis et al. [6] found evidence that being known to other members (i.e., social embeddedness) and active (i.e., fecundity) increase the likelihood of songs being *remixed* (i.e., reused, in case of audio/video products). Stanko [26] investigated why some 3D-printable objects in the Thingiverse creative community are more successful than others and found that their degree of generativity is proportional to the degree of interaction of the authors with other community members. Similar findings are reported by Settles and Dow [22] who analyzed the music community of FAWM and found that the prior exchange of direct messages between musicians is the factor that contributes the most to the successful completion of their collaborations. Hill and Monroy-Hernández [13] performed a study on the success of video animations in Scratch and found that the likelihood of engendering derivative works is related to cumulativeness (i.e., remixes themselves are reused more than *de novo* content), which is in direct contrast with the finding by Cheliotis et al. [6] on the degree of derivativity (i.e., the 'newer' the content, the higher the likelihood of remix). The contrasting findings from these two works provide additional motivation to further study the antecedents of artifact reuse in collaborative songwriting communities.

The interpretation of successful collaborations in OSS communities follows the one given above for online arts communities. Rather than evaluating the success of a piece of software developed by an OSS community from the perspective of its end-users (e.g., through the number of downloads), we keep the perspective of the creators (i.e., the developers) who sustain the community by writing software together. Accordingly, here we look at success within OSS communities in terms of code contributions (i.e., fixes, new features), which are merged into a project repository to improve the product. The review of prior works reveals that the success factors influencing the acceptance of these contributions are both social and technical in nature [12, 28]. Ducheneaut [9] found that changes coming from submitters who are already known to the core development team have higher chances of being accepted. In fact, the rise of 'transparent' social coding platforms (e.g., Bitbucket and GitHub) has made social factors more prominent, as code reviewers have started to make inferences about the quality of contributions not only by looking at code quality but also using social aspects like developers' reputation (e.g., number of followers) as auxiliary indicators [7, 18]. Tsay et al. [28] found that contributors' reputation in GitHub (i.e., track record) and their social closeness to the development team are stronger antecedents of contribution acceptance than technical factors (e.g., the inclusion of test cases).

Consistently, in another study on GitHub, Gousios et al. [11] found that only ~13% of contributions are rejected for purely technical reasons.

Previous work has overlooked to compare success factors across these types of communities. One notable exception is the work of Luther et al. [16], which identified frequent communication, project leader's experience, and reputation as common factors reported in the related literature. Our research aims at furthering this comparison.

To this end, we perform an extension of our previous study [3], where we tested which factors from prior work in both OSS and art communities are predictive of successful collaboration in Songtree, a songwriting community [5]. To achieve better generalizability of findings, here we replicate the study on two more online communities for collaborative music creation: ccMixter and Splice. From the original study, we inherit a set of five hypotheses. The subsequent regression analysis and measures are adapted to accommodate the differences in the new datasets. Furthermore, in this work, we complement the results from the quantitative analysis with the analysis of a questionnaire intended to collect further insights into the factors triggering song reuse as well as feedback on collaborative features.

The main result from this study is the evidence that songs with positive feedback and a low degree of derivativity, as well as those by authors with a high status and an easy to spot profile, are associated with higher odds of being reused. Based on the preliminary results from the analysis of the questionnaires, we also provide suggestions that may be potentially useful to the designers of online community platforms.

The remainder of the paper is organized as follows. In Section 2, we describe the study and its hypotheses. The experimental datasets and the features extracted are presented in Section 3. Results from our research methods are reported in Section 4 and discussed in Section 5. Finally, conclusions are drawn in Section 6.

## 2 THE STUDY

To perform the extension of the original study on Songtree, we extracted the content from two other creative music communities, Splice, and ccMixter. Below, we first describe the three communities, highlighting their similarities and differences, along with the specific mechanism of song reuse in each of them. Then, we discuss the set of five hypotheses tested.

### 2.1 Communities Under Study

ccMixter[2] is an online community born in 2004 where artists share music released under Creative Commons (CC) licenses. As such, members can freely (and legally) reuse any song to create their own remixes. To keeps track of content reuse, authors are requested to declare whether the uploaded content is a *de novo* contribution or a remix, in which case the reused sources must be reported. Community members can recommend songs (i.e., like), write reviews (i.e., comment), and create playlists of their favorite music. They also have a profile page with their image, short bio, and contacts; information about their recommendations, reviews, uploads, and playlists are available too.

Splice[3] is an online platform launched in 2014, which offers cloud storage to upload music samples and several plugins (both free and paid) for music recording and track mixing. Musicians create music either starting new projects or by *splicing* existing ones, that is, creating in their personal space copies that can be freely modified. Other members can be added to projects to collaborate. Private messages can be used to coordinate the work. Users are allowed to comment and like content. They also have a personal page where their bio and statistics of their activity are shown (e.g., tracks they played and liked, followers and followings).

[2] http://ccmixter.org/view/media
[3] https://splice.com/explore

Songtree[4] is a collaborative songwriting community launched in 2011, where users *overdub* existing songs by mixing (i.e., adding) one additional track. Songtree leverages the metaphor of collaborative songwriting as a tree where each new song uploaded is the root and the songs derived from it, through overdubbing, are the nodes that branch out. Over time, a song tree grows as new overdubs are derived from any of the songs in it, with leaf songs representing alternative evolutions of the same root song.

Hereinafter, we will use the more general term *reuse* to refer to the derivation of content from existing music in any of the three communities analyzed. A generic song generated through the reuse of preexisting content will be referred to as a *derived song*. Instead, we will only use the terms overdubs, splices, and remixes when discussing individually the specifics of the related community.

Albeit similar, the three communities exhibit some differences that are worth noting. These differences concern (i) the purpose of the community, (ii) the presence of features to support coordination, and (iii) the underlying reuse mechanism. As for the purpose, the main goal of ccMixter is to spread the CC licenses for making music that is freely and legally reusable. Instead, albeit the content available on Splice and Songtree is also shared under a CC license, the companies behind the communities have a commercial interest. Splice offers to its members a rent-to-own pricing model for software synthesizers. Songtree aims at spreading the use of n-Track, the commercial digital-audio-workstation software produced by the same parent company.

Regarding coordination, Splice is the only platform that allows its users to create projects, each having members explicitly assigned to it and a repository for managing artifacts. On the contrary, coordination in ccMixter and Songtree appears to be rather *ad hoc*.

Finally, regarding the mechanism of reuse, ccMixter and Splice are more similar because they allow the reuse of multiple artifacts to create a new one (i.e., many-to-one reuse), as opposed to Songtree, which only allows users reusing one existing song by adding an extra track (i.e., one-to-one reuse).

## 2.2 Hypotheses

We inherit five hypotheses from the original study [3]. The hypotheses were designed to study the antecedents of *overdubbing* (now *reuse*) on two different levels of analysis related to the artifacts themselves (i.e., the songs, see H1-H3) and the authors (see H4 and H5). The hypotheses are listed below, slightly modified to comply with the 'agnostic' terminology defined earlier.

As for the song-related hypotheses, both OSS and artistic communities leverage several social network-like features such as the ability to provide feedback (e.g., likes, comments). As such, we argue that those generating more reactions are more popular and, thus, more likely to be reused.

*H1: Songs that generate a high amount of reactions are more likely to be reused.*

Also, we hypothesize that rather than digging up old content, community members might be more prone to reuse songs recently uploaded.

*H2: More recently uploaded songs are more likely to be reused.*

Finally, due to the contrasting findings of prior research, we define the third hypothesis to gather further evidence on the relationship between the degrees of generativity and derivativity in creative communities [6, 14]. In the original study on Songtree, we hypothesized that more 'mature songs' – i.e., derivations of derivations, closer to leaf nodes rather than the root – would generate fewer overdubs because they leave less space for extension. Hence, the third hypothesis, updated to the current terminology, reads as follows:

[4] http://songtr.ee

*H3: Derived songs are less likely to be further reused.*

Regarding author-related hypotheses, prior research on both OSS (e.g., [28]) and arts communities (e.g., [6, 16]) highlighted that authors' prominence and their social embeddedness are antecedents of success. Hence, we expect that content generated by community members with a high status is reused more often.

*H4: Songs by authors with a high status in the community are more likely to be reused.*

Finally, previous research on collaboration in OSS communities has found that the identity of the submitters is a very significant proxy for the quality of contributions [12]. Besides, Luther et al. [16, 17] found that being able to browse members' history of contributions increases the chance of successful collaboration. Community websites allow artists to maintain a profile page providing an avatar picture, a short bio, and personal contact information. Some members may perceive that the effort put into curating their personal space may reflect the attention put into creating their music. As such, we hypothesize that community members are more inclined to reuse songs by fellow authors whose profile is more easily recognized from their customized avatars.

*H5: Songs by authors with a customized avatar are more likely to be reused.*

## 3 RESEARCH METHODS

To perform our investigation, we use both quantitative and qualitative methods. First, to test our hypotheses we build the three experimental datasets from Songtree, ccMixter, and Splice. Using the features extracted from such datasets, we fit a logistic regression model to associate the defined features (i.e., the predictors) measured in each community with the likelihood of a song being reused. Then, a short user survey is administered, to garner additional qualitative insight into community members' perceptions of further factors triggering song reuse as well as to collect feedback on collaboration features, missing and already implemented, in the songwriting platforms.

### 3.1 Datasets

A breakdown of the original dataset from Songtree and the two new datasets from ccMixter and Splice is available in Table 1. The datasets contain information about authors and songs, both new and derived. The breakdown shows that Songtree and ccMixter are, size-wise, quite similar communities and both much larger than Splice, which is the most recent one, launched about 4 years ago.

In the original study, the Songtree dataset was built from the entire dump of the database provided by the community administrators in December 2016. As initial preprocessing steps, we first removed all the songs and their overdubs created prior to March 2015, i.e., when administrators were still participating actively to kick-start the community. Then, we filtered out all the new songs and their overdub created in the last 27 days before the dump. This 27-day threshold corresponds to the 90th percentile of the overdub time intervals between the upload of a song and that of its first overdub. In other words, 90% of the songs in the rather sparse dataset have received their first overdub within 27 days since their upload. This step was necessary to avoid *right censoring* issues [1] and ensure that there was sufficient time to observe the event of interest (i.e., receive at least one overdub) for all the selected songs. We also removed contest songs – i.e., uploaded by the Songtree team to start contests with prizes awarded to the best overdubs – as well as hidden and closed songs, which are impossible to discover or derive, respectively. As for users, we filtered all the non-authors and administrators account. At the end of the preprocessing, we ended up with a final dataset of 16,998 songs (10,672 new songs plus 6,326 derived/overdubs), and 3,790 authors (see Table 1).

Table 1. A breakdown of the final datasets extracted from the three communities. *New songs* are those created *de novo* and uploaded to the platform, *Derived songs* refers to songs that are created by reusing an existing one, and *Reused songs* are those that have generated at least one derived song.

| | **Songtree** | **Splice** | **ccMixter** |
|---|---|---|---|
| **Authors** | 3,790 | 284 | 3,327 |
| **Songs** | 16,998 | 1,168 | 27,013 |
| **New songs** (% of Songs) | 10,672 (63%) | 983 (84%) | 9,463 (35%) |
| **Derived songs** (% of Songs) | 6,326 (37%) | 185 (16%) | 17,550 (65%) |
| **Reused songs** (% of Songs) | 2,638 (15%) | 116 (10%) | 10,656 (39%) |
| **New songs reused** (% of New songs) | 1,739 (16%) | 105 (10%) | 5,660 (60%) |
| **Derived song reused** (% of Derived songs) | 899 (14%) | 11 (5%) | 4,996 (28%) |

In the case of ccMixter and Splice, we were unable to get access to the database dumps. We, therefore, had to resort to retrieving the content (up until February 2018) by developing two custom web scrapers in Python, using the Scrapy and Selenium libraries. Following the same steps used before for Songtree, we preprocess both datasets to remove non-authors and contest songs, and filter recently uploaded songs to avoid censoring issues. As shown in Table 1, we end up with 1,168 songs (983 new songs plus 185 derived/splices) and 284 authors in the case of Splice; for ccMixter, we end up with 27,013 songs (9,463 new songs plus 17,550 derived/remixes) and 3,327 authors.

The data in Table 1 allow us to draw a clearer picture of the reuse habits within each of the three communities. First, we observe that in both Songtree and Splice *New songs* (i.e., those created *de novo*) outnumber *Derived songs* (i.e., those created by reusing another song); the opposite is observed in ccMixter, where 65% of the songs in the dataset comprises *Derived songs*. Consistently, the percentage of *Reused songs* (i.e., songs that have generated one or more derived songs) in ccMixter (39%) is larger than Songtree (15%) and Splice (10%). Second, we note that *New songs* appear to be more generative, especially in ccMixter where 60% of *New songs* are reused compared to only 28% of *Derived songs* being further reused. This difference is less noticeable in Songtree (16% vs. 14%) and Splice (10% vs. 5%).

To facilitate independent replications of our study, we have made the ccMixter and Splice datasets available on GitHub.[5] The Songtree dataset is not included because we have signed a confidentiality agreement to ensure the non-disclosure of the content.

### 3.2 Features

From the three datasets, we extract several features to inform our analysis. For the sake of readability, they are reported concisely in
Table 2.

The outcome of the statistical model to develop is whether a song has been reused (i.e., overdubbed/remixed/spliced) or not. Thus, as in the original study, we define a dichotomous dependent variable, here named `reused`.

[5] https://github.com/collab-uniba/music_scrapers

Table 2. The features extracted from each dataset in this study.

| | Feature | Community | | | Feature Type | Feature Description | H |
|---|---|---|---|---|---|---|---|
| | | Songtree | Splice | ccMixter | | | |
| | `reused` | ✓ | ✓ | ✓ | nominal | Whether the song has received any overdub/remix/splice or not. Values: {*Yes, No*} | - |
| Song level | `#likes` | ✓ | ✓ | ✓ | ratio | No. of likes that the song received by community members | H1 |
| | `#comments` | | ✓ | ✓ | ratio | No. of comments that the song received by community members | |
| | `#bookmarks` | ✓ | | | ratio | The number of times that the song has been bookmarked | |
| | `time_since_upload` | ✓ | ✓ | ✓ | interval | Time (in days) until the first derived song is uploaded | H2 |
| | `song_depth` | ✓ | | | ratio | The distance in number of nodes from the root song that started the song tree. It is 0 for root songs | H3 |
| | `is_derived` | | ✓ | ✓ | nominal | Whether the song is derived from another or new. Values: {Yes, No} | |
| Author level | `#followers` | ✓ | ✓ | | ratio | No. of users following author's activities | H4 |
| | `author_ranking` | ✓ | ✓ | ✓ | ratio | The measure of author embeddedness in the community divided by the sum of the number of artifacts uploaded and the time since the author joined the community | |
| | `has_avatar` | ✓ | | ✓ | nominal | Whether the author has uploaded a profile picture or not. Values: {*Yes, No*} | H5 |

As for the independent variables (i.e., the predictors in the model), we reuse the same features defined in the original study wherever possible. In fact, the differences in the communities under study and their datasets force us to make some modifications compared to the original study. To ease the interpretability of the findings, we detail these adaptations below.

*H1 – Reactions.* Regarding the first hypothesis, we quantify the amount of feedback generated by songs in terms of the number of likes, comments, and bookmarks. The feature `#likes` is available in the three communities. On the contrary, the feature `#bookmarks` is only available in Songtree and `#comments` only in ccMixter and Splice.

*H2 – Recent songs.* The second hypothesis is operationalized by computing the feature `time_since_upload`, which measures the time interval in days between the upload of a given new song and that of the first song derived from it.

*H3 – Derived songs.* To operationalize the third hypothesis about the degree of generativity of new vs. derived songs, in the original study we leveraged the song-tree data structure to measure the `song_depth`, that is, the length of the path in the tree from the root node (representing a new song) to any other node in the tree (representing a derived song). This measure allowed us to capture the degree of derivativity of a song and also the amount of work that has been already done on it. However, songs are not arranged in trees in either Splice or ccMixter. Thus, in this study, we are forced to extract an alternative and simplified feature, `is_derived`, defined as a dichotomous variable indicating whether a song is new or reused.

*H4 – Status in the community.* To quantify the status of authors in the community, we first extract the `#followers`. The feature is only available in Songtree and Splice, but not in ccMixter, which is the less social network-like community of the three. Then, we define and extract the feature `author_ranking`, which allows determining an order on the level of embeddedness and fecundity of authors in their community. For each community, the embeddedness and fecundity metrics vary because they are defined in terms of the different features available therein, as reported in Table 3. For example, in the case of Splice, embeddedness is quantified by including the number of likes, plays, and followers. Author's fecundity is computed in terms of the number of songs shared in the community plus either tenure, in the case of Songtree and Splice, or the number of songs reviewed, for ccMixter. Finally, `author_ranking` is obtained by normalizing the embeddedness by author fecundity.

*H5 – Recognizable avatar.* The fifth hypothesis is operationalized by extracting a dichotomous feature (`has_avatar`) that captures whether an author has customized the profile picture or rather kept the predefined one assigned at registration time.

In Table 4, we report some descriptive statistics of the features defined above, as extracted from each dataset. Overall, the table shows that not all features are used similarly across the three communities. For example, despite the difference in size, the average number of comments per song is similar in ccMixter and Splice. Similarly, we note that most of the authors customize their avatar in both Songtree and ccMixter. On the contrary, we note that liking songs and following other members is much more widespread in Splice than in the other two communities. As for time, the mean interval for songs to be reused is shorter in Songtree (171 days) than in Splice (630) and ccMixter (2,515).

Finally, we note that the three datasets are cross-sectional, i.e., they provide a snapshot of multiple data collected at one point in time. Cross-sectional data are typically sensitive to *reverse causality*[6] problems. In the case of Songtree, the availability of the official database dump (i.e., the entire history of events recorded in the community) enabled us to compute the value of the features at a point in time *just prior* to the event of interest '*song has been derived*' (e.g., the number of likes received by a song `S` *right before* it was overdubbed by another song `T`). As such, the measures extracted from Songtree mitigate reverse causality issues and allow us to make inferences about the underlying *direction of causality* between the observed likelihood of song being reused and the occurrence of any of the predictors. Instead, because we retrieved the content of both ccMixter and Splice through web scraping, we do not have access to the history of events.

As such, for these two communities, the extracted features are subject to reverse causality issues, that is, we only know how many likes a song `S` has received at the time of data collection but not at the time when it was reused for the first time. In Section 4 and 5, we discuss respectively the validity and the implications of the findings due to such reverse causality concerns.

Table 3. Embeddedness and fecundity measures defined for each community.

| | **Embeddedness** | **Fecundity** |
|---|---|---|
| Songtree | $\#followers + \#likes + \#plays + \#derived_plays$ | $\#shared_songs + tenure$ |
| Splice | $\#followers + \#likes + \#plays$ | $\#shared_songs + tenure$ |
| ccMixter | $\#song_reviews_received + \#remix + \#playlisted$ | $\#shared_songs + \#song_reviews_written$ |

[6] Reverse causality refers to either a direction of cause-effect contrary to expectation or a two-way causal relationship between the predictors and the dependent variable.

Table 4. Descriptive statistics for the features extracted from each dataset in this study.

| Feature | | Community | | |
|---|---|---|---|---|
| | | Songtree (n=16,998) | Splice (n=1,168) | ccMixter (n=27,013) |
| #likes | min | 0 | 0 | 0 |
| | median | 1 | 4 | 6 |
| | mean | 1.7 | 22 | 8.5 |
| | max | 59 | 1813 | 235 |
| | total | 29,435 | 25,980 | 229,879 |
| #comments | min | - | 0 | 0 |
| | median | - | 0 | 2 |
| | mean | - | 3 | 2.8 |
| | max | - | 368 | 45 |
| | total | - | 3,269 | 75,299 |
| #bookmarks | min | 0 | - | - |
| | median | 0 | - | - |
| | mean | 0.22 | - | - |
| | max | 14 | - | - |
| | total | 3,705 | - | - |
| time_since_upload | min | 12 | 9 | 10 |
| | median | 142 | 739 | 2531 |
| | mean | 171 | 630 | 2515 |
| | max | 590 | 1469 | 4856 |
| song_depth | min | 0 | - | - |
| | median | 0 | - | - |
| | mean | 0.62 | - | - |
| | max | 16 | - | - |
| is_derived | yes | - | 185 | 17,550 |
| | no | - | 983 | 9,463 |
| #followers | min | 0 | 0 | - |
| | median | 5 | 33 | - |
| | mean | 26.9 | 150 | - |
| | max | 145 | 5304 | - |
| | total | 456,644 | 174,844 | - |
| author_ranking | min | 0 | 0 | 0 |
| | median | 13 | 10 | 0.9 |
| | mean | 26 | 35 | 1.1 |
| | max | 3408 | 846 | 41 |
| has_avatar | yes | 13,390 | - | 22,965 |
| | no | 3,608 | - | 4,048 |

### 3.3 User Survey

Although we reached out to the administrators of all three platforms to propose a user survey, we got a positive response only from Songtree. We managed to include a couple of both closed questions and open-ended ones as part of a broader-scope questionnaire administered by Songtree administrators. The closed questions were intended to assess the participants' level of experience with the platform and their role in the community (i.e., songwriters vs. listeners). As for the open-ended questions, we asked users about (i) the perceived factors that may influence their decision to reuse a song and (ii) comments and/or features requests concerning collaboration within the platform.

The questionnaire was made available in the period between Jan. and Apr. 2018, during which we received 97 valid responses. Over a half of the survey participants report experience with the platform of more than 6 months (n=52, see Figure 1a), and 60% of them has uploaded at least 1 overdub since joining (n=57, see Figure 1b). As for the open-ended questions, because they are specifically intended for Songtree authors who may have experience with overdubbing and collaborating with other fellow musicians, all the responses from non-authors with no overdubs and less than 6 months of experience are discarded for this analysis. Nonetheless, ignored responses have been double-checked for the presence of valuable insights, though they are either blank or contain general appreciation for the platform. Eventually, we retain 49 valid responses, which are analyzed to identify patterns and common themes.

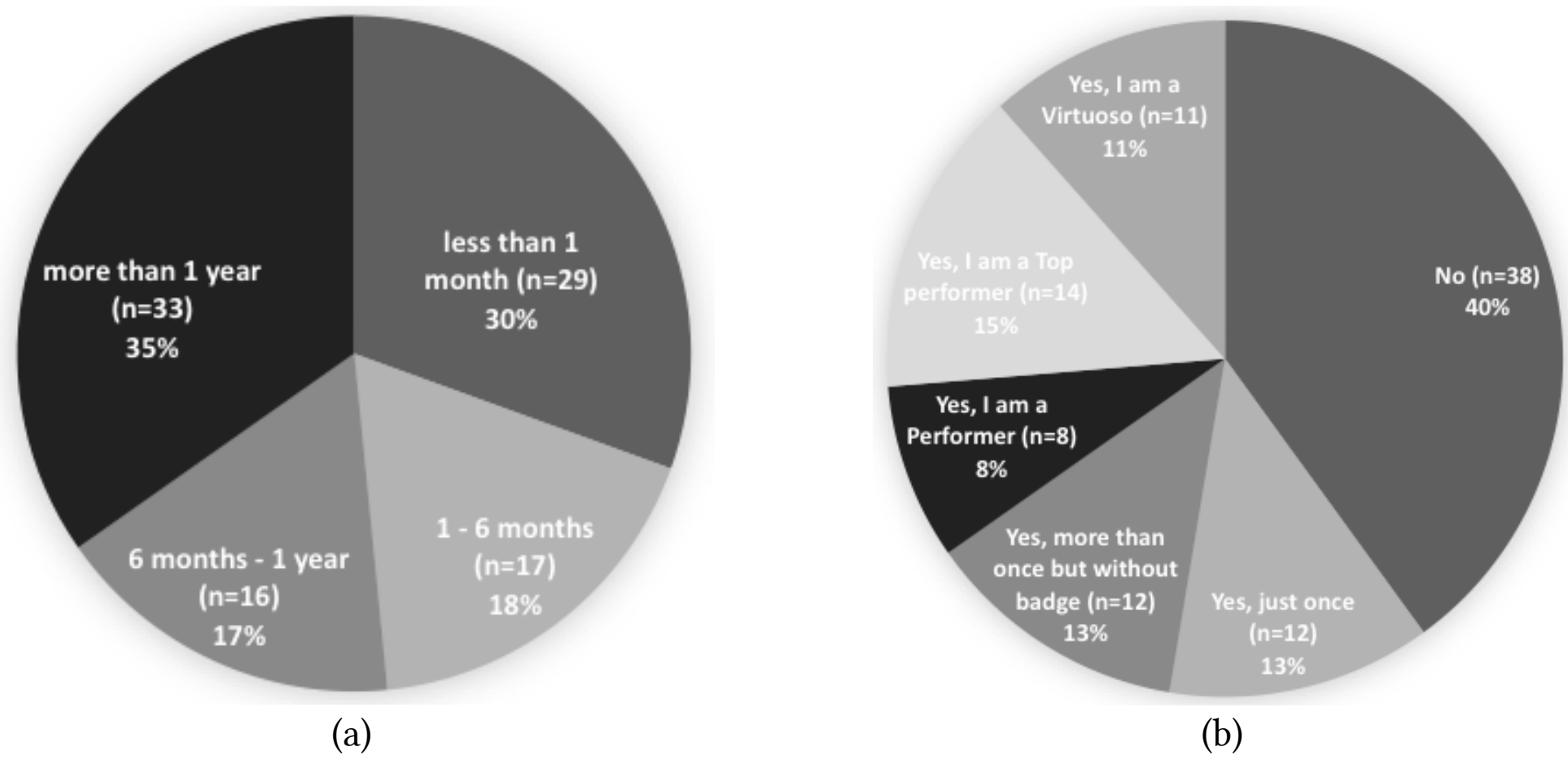

Figure 1. Breakdown of survey respondents' (a) experience with Songtree and (b) number of overdubs created (note: performer=10-39 overdubs, top-performer=40-199, virtuoso=200+).

## 4 RESULTS

### 4.1 Quantitative Analysis

In this section, we examine our hypotheses and how each feature (i.e., predictor variable) is associated with our dependent variable that models a successful collaboration, i.e., whether a song has been reused. Specifically, for each community, we fit a logistic regression model to the extracted data for predicting the binary outcome variable `reused`.

To perform the logistic regressions, we use the `glm` function from the `stats` R package. The continuous variables of type ratio, which follow a long-tail distribution, are log-transformed to reduce the skewness. To further ease the assessment of the relative importance of the predictors, each of the continuous variables in the model is standardized, such that the mean of each measure is 0 and the standard deviation is 1. Furthermore, to ensure compliance with the logistic regression assumption, we check the ccMixter and Splice datasets for multicollinearity problems [19], which have been already analyzed and fixed for Songtree in the original study. Thus, we first compute the correlation matrix for the variables in each model and remove the features that have strong pairwise correlations (i.e., ≥ 0.7). In particular, for ccMixter, we remove the number of songs uploaded that is highly correlated with the feature `#comments`, which we retain instead. For Splice, we remove the number of plays and splices received by a song, as they are both highly correlated with each other and with the feature `#likes`, which we retain instead. Then, after fitting the models, we use the Variance Inflation Factor (VIF) to check again for multicollinearity. All VIF values are smaller than 3. Accordingly, we exclude violations of the multicollinearity assumption.

In Table 5, we report the results related to four logistic models, one for each of the three studied communities plus a fourth model fit to another dataset, SongtreeRC, built from the Songtree database by intentionally extracting the features of `author_ranking`, `#followers`,

`time_since_upload`, `#bookmarks`, `#comments`, and `#likes` at the time of data collection (i.e., subject to reverse causality issues), as in the case of ccMixter and Splice. This way, through the performance of this fourth model we can further assess the robustness of the Songtree model, for which the same continuous features are properly extracted just before the event of interest. We note that while the definitions of these independent variables are the same in the Songtree and SongtreeRC datasets, they are not correlated. Instead, the outcome variable `reused` in the two datasets is exactly the same.

Except for Splice, all predictors in the other models are statistically significant at the 1% level (i.e., $p < 0.01$) or smaller. In the case of Splice, instead, predictors are significant the 1%-5% level, except for `#followers` and `author_ranking`, which are not significant. The significance of terms was obtained from the Wald test in the `ANOVA`, as implemented by the `car` package. Given the large size of our datasets, however, here we discuss predictor contributions in terms of odds ratio, which is an unstandardized effect size statistic that tells the direction and the strength of the relationship between predictors and the odds that a song is reused, i.e., the increase or decrease of the odds of ‘success’ occurring per ‘unit’ of the measure. We remind the reader that an odds ratio close to 1 means that exposure to feature A (i.e., one of the considered predictors) does not affect the odds of a song being reused. An odds ratio far smaller than 1 is instead significantly associated with lower odds. Conversely, an odds ratio much larger than 1 means that there are higher odds for a song to be reused with exposure to predictor A.

To evaluate the goodness of fit of each model, in Table 5 we report the Nagelkerke’s pseudo $R^2$, a statistical measure that represents the percentage of the response variable variation that is explained by a model. Results show that our Songtree model fits the data very well, as it explains ~90% of the variability of the data ($R^2 = 0.89$). The fit is largely worse in the case of SongtreeRC, as shown by the smaller $R^2$ (0.31, ~35% less variability explained). A similar fit is obtained with the ccMixter model ($R^2$=0.33). Similarly, the Splice model does not fit the data very well as it is able to explain only about 20% of the variability in the data ($R^2$=0.19).

Table 5. The logistic models of the likelihood of song reuse. Predictors with a large effect size (odds ratio) are shown in bold.

| | **Hypothesis** | **Feature (predictor)** | **Odds Ratio (OR)** | | | |
|---|---|---|---|---|---|---|
| | | | *Songtree* | *SongtreeRC* | *Splice* | *ccMixter* |
| **Song** | | (Intercept) | 0.07 | 0.07 | 0.09 | 1.98 |
| | Reactions (H1) | `#likes` | 1.18$^{**}$ | **1.52**$^{***}$ | **1.54**$^{**}$ | **2.80**$^{***}$ |
| | | `#comments` | - | - | **1.44**$^{**}$ | 0.93$^{**}$ |
| | | `#bookmarks` | **2.12**$^{***}$ | **1.27**$^{***}$ | - | - |
| | Recent Songs (H2) | `time_since_upload` | **0.02**$^{***}$ | **0.72**$^{***}$ | **1.43**$^{*}$ | **1.34** $^{***}$ |
| | Derived Songs (H3) | `song_depth` | **0.56**$^{***}$ | **0.73**$^{***}$ | - | - |
| | | `is_derived` (default: `false`) | - | - | **0.47**$^{*}$ | **0.11**$^{***}$ |
| **Author** | Status in Community (H4) | `#followers` | 0.82$^{***}$ | **1.42**$^{***}$ | 1.17 | - |
| | | `author_ranking` | **9.48**$^{***}$ | **2.32**$^{***}$ | 1.13 | **1.36**$^{***}$ |
| | Customized Avatar (H5) | `has_avatar` (default: `false`) | **3.11**$^{***}$ | **1.47**$^{***}$ | - | **1.26**$^{***}$ |
| **Nagelkerke’s pseudo $R^2$** | | | 0.89 | 0.31 | 0.19 | 0.33 |
| **AUC** | | | 0.98 | 0.83 | 0.66 | 0.79 |

sig.: ‘***’ p<0.001; ‘**’ p<0.01; ‘*’ p<0.05

To assess the predictive capability of the models, we compute the Area Under the Receiver Operating Characteristic (AUC) curve. A ROC curve plots the performance of a binary prediction model as the trade-off between its ability to recall the positive instances in a dataset (i.e., the true positive rate, or how many songs predicted as reused have been actually reused) and the false positive rate (i.e., how many songs are misclassified as having been overdubbed). We split the dataset into training (70%) and test (30%) sets, using the stratified sampling function of the `caret` package to maintain the same proportion of `reused` songs across the two subsets. Because the AUC performance of a random baseline classifier is 0.5 and the performance of our models range between 0.66 (Splice) and 0.98 (Songtree), we conclude that all our models outperform random guessing, even more so in the case of Songtree and ccMixter.

Below, we report the findings from our quantitative analysis with respect to each of the hypotheses defined.

**H1 – Songs that generate a high amount of reactions are more likely to be reused.**

To test this hypothesis, we examine the association between the probability of a song to be reused and the number of likes, bookmarks, and comments received.

In the original study on Songtree, we found general support for the hypothesis that songs that receive more likes and bookmarks have higher odds of being overdubbed, although the effect of `#likes` (OR=1.18) is small compared to the other predictor `#bookmarks` (OR=2.12). Similarly, for ccMixter, we observe a strong and positive effect of the `#likes` (OR=2.80) while that of `#comments` is negligible (OR=0.93). Finally, in the case of Splice, H1 is supported as both `#comments` (OR=1.44) and `#likes` (OR=1.54) are strong and positively associated with higher likelihood of reuse.

**H2 – More recently uploaded songs are more likely to be reused.**

The test of the second hypothesis on the effect of time interval between a parent song and the derived one reveals mixed results. In the original study, we found strong support for H2, as `time_since_upload` is the strongest predictor in the model (OR=0.02), indicating that older songs have very low odds of being derived. On the contrary, H2 is not supported in the case of Splice and ccMixter, where songs keep on being derived over time (OR=1.43 and 1.34 for `is_derived`, respectively).

**H3 – Derived songs are less likely to be further reused**

With H3, we tested the effect of the degree of derivativity of songs on their capacity to generate other derived songs. We observe strong support for this hypothesis in the original study on Songtree (`song_depth` OR=0.56) as well as in this study on Splice (OR=0.47) and ccMixter (OR=0.11), despite the use of a simplified predictor (`is_derived` instead of `song_depth`).

**H4 – Songs by authors with a high status in the community are more likely to be reused**

We find mixed support for our fourth hypothesis. In the original study, we found partial support that having a higher status in the Songtree community will increase the odds of song overdub. In fact, while the `author_ranking` odds ratio is positive and strong (9.48), the `#followers` predictor has an unexpected negative effect on the odds of reuse, albeit its effect size is small (OR=0.82). In Splice, H4 is not supported since both of the predictors are not significant. Finally, in ccMixter the `#followers` predictor is unavailable, but `author_ranking` is positively associated with higher odds of song remix (OR=1.36).

**H5 – Songs by authors with customized avatars are more likely to be reused**

Finally, regarding our fifth hypothesis that songs by authors whose profile is easier to spot, we find support for both Songtree (`has_avatar` OR=3.11) and ccMixter (OR=1.26). However, we are unable to test the hypothesis for Splice.

#### Limitations

We identified a few limitations that affect the construct validity of the findings from this study. Construct validity concerns the degree of accuracy to which the variables (i.e., features) measure the constructs of interests.

In this study, we have collected cross-sectional data, which do not allow to clear the causality nexus. However, in the case of Songtree, the availability of the entire dataset dump gave us access to the entire history of events and, thus, made it possible to mitigate reverse causality issues by extracting the predictors just before the event of interest is observed. Therefore, for Songtree we can make inferences about the underlying *direction of causality* between the observed likelihood of song being reused and the occurrence of any of the predictors in the logistic model. On the contrary, for Splice and ccMixter the history of events is not available because data have been collected via web scraping. Due to this limitation, in the analysis of these two communities, we can only identify positive/negative relationships between the significant predictors and the outcome variable. In future work, we will collect further snapshots of Splice and ccMixter and perform longitudinal analyses that will allow us to make inferences about causality.

Due to differences in the communities and their datasets, we were not able to extract the number of comments on Songtree (H1) and we had to 'flatten' the ratio predictor `song_depth`, measuring the degree of derivativity of a song as its distance from the root node in the song tree, to the dichotomous variable `is_derived`, capturing only whether a song is a root (i.e., new) or not (i.e., derived). Likewise, the `author_ranking` construct varies across the three communities and, therefore, we cannot exclude that the different operationalizations may account for some of the differences observed in the logistic models.

Regarding the `has_avatar` construct, which we defined as a proxy for author profiles that are easier to recognize, we acknowledge that users may unintentionally select custom profile images that are hard to recognize nevertheless.

### 4.2 Qualitative Analysis

Complimentary to the quantitative analysis of success factors, we asked the survey respondents (i) what other factors they perceive as triggering their will to start collaborating with others by overdubbing and (ii) what comments/feature requests they had about collaboration within the Songtree platform.

To analyze the open-ended questions, we applied hybrid card sorting [25]. First, one of the authors reported the answers in an Excel spreadsheet and started to identify common themes. Then, another author received the spreadsheet with the suggested categories and performed the card sorting individually. Finally, a joint session was held to clarify ambiguities and update the categories until a mutual agreement was reached. The final categories resulting from the analysis are reported in Table 6.

As for the first question, we find 6 common themes that influence their decision to overdub. We note that we ignore all the comments that mentioned high recording quality as a reason for picking a song to overdub because it seems rather a technical requirement (e.g., "*only songs that sound clean or well produced*"). Regarding the responses to the second question, we also identify 6 categories but, in this case, we consider off-topic and, therefore, ignore all comments about features completely unrelated to collaborative songwriting (e.g., "*Please, improve the playlist editor*"). Interestingly, the categories 1-5 identified from the analysis of the responses to the first question can be directly mapped onto those identified for the other one (see Table 6).

Table 6. The categories resulting from the analysis of the two open-ended questions in the user survey (the numbers within parenthesis represent the coverage of responses within the category).

| Category | Triggers of overdubbing | | Comments / feature requests about collaboration | |
|---|---|---|---|---|
| 1 | Genre (2) | | Flexible genre categorization (2) | |
| 2 | Friends network | Invitations and @mentions (3) | Friends network | Virtual bands and virtual albums (2) |
| 3 | | Followings (2) | | Music producers (1) |
| 4 | | Opportunity for contributing (2) | | Better chat (1) |
| 5 | Contests (3) | | Songtree awards (2) | |
| 6 | Personal taste (8) | | Abolish author ranking (2) | |

Regarding the first question, the most common trigger of overdubbing is **personal taste** in music (e.g., "*I overdub what I like!*", "*Music that inspires me*", "*Whatever catches me*"). A couple of respondents report choosing a song to overdub by browsing the music by **genre**. As such, it is not surprising that those respondents also ask for the adoption of a **more flexible genre categorization** approach to organize and browse songs, since the currently adopted top-down taxonomy imposes categories that seem to be too broad (e.g., "*Especially being a blues player, I seek other blues lovers [but] it seems everything is labeled blues while it really could be anything*").

Other factors triggering collaborations are related to the activities of the **network of friends** in Songtree (e.g., "*My main trigger for overdubs is 'My Feed' updates: I see songs performed by the friends that I am following, and [...] I am in!*"). Several respondents report to start an overdub when they are explicitly invited by people they have befriended (e.g., "*Mainly [I overdub] because I am invited*") or when they are mentioned in comments by friends who think they can contribute to that song (e.g., "*when others tag me or mention me in a comment and [make me] discover new interesting things*", "*they [mention] me when songs need bass*"). Consistently, a couple of respondents provide extremely detailed feedback about collaborative features that Songtree should implement to further support collaboration along the entire lifecycle of album creation. Specifically, they call for adding support for 'virtual bands' formation ("*Let's say that there is a song idea. Authors can invite certain musicians to work on this idea together - set a "virtual band" that will work on this project*") with overdubs displayed together in one multitrack interface. These 'virtual bands' should be able to create and manage entire 'virtual albums' that "*group their songs together*" and export their work also to services like YouTube, where they could further promote their music. Besides musicians and singers, the platform should add support also for those acting as 'producers.' Finally, one respondent adds that the chat service should be overhauled because "*[making] the chat more interactive [...] make collaboration around songs easier and more interesting.*"

In Songtree, platform administrators often launch contests to award with prizes the best overdubs of a contest root song. Thus, it is not surprising that three respondents mentioned **contests** as one of the main factors driving the decision to overdub. Yet, contests are competitive rather than collaborative efforts. A couple of respondents suggest that contests should become **Songtree awards**, that is more varied events organized on regular basis to award "*the best newcomer, the best collab etc., all voted by us.*"

While many of the responses to the second questions focus on requesting new features, two users strongly suggest the **abolishment of author ranking**, a feature that is at the very core of the current platform implementation. According to them, author ranking fosters competition rather collaboration, which is perceived as the ultimate goal of the platform ("*There's a lot to love about [Songtree] because I can play with other musicians from around the world. But the competition [...] for the ranking is all over the place and overwhelming. It creates rivalry and bad ambiance among*

*Songtree users.*", "*The ratings [...] I deem somewhat counterproductive, make people compare, be unhappy or feel misjudged... jealous!*").

**Limitations**

Because we were only able to administer the questionnaire only to Songtree users, the observations derived in the section above may not apply to ccMixter and/or Splice. Hence, the results of the qualitative analysis performed in this paper have limited generalizability.

## 5 DISCUSSION

In this section, we discuss the findings from the current extension, comparing them to both the original study and prior research, while also speculating on the motivations of some contrasting results. We build on the findings from quantitative analysis to provide actionable insights to authors seeking to increase their status in the studied communities. Furthermore, we use the findings from the qualitative analysis of the survey to explain some of the results of the regression analysis. We also discuss the implications for researchers and speculate about the findings to provide suggestions potentially useful to the designers of platforms for online communities. Finally, we discuss the limitations of the current study and derive new research questions for our future work.

### 5.1 Likelihood of Song Reuse

Overall, the results of the quantitative analysis provide general support for our hypotheses. While the detailed discussion about results for each of the hypotheses is provided below, here we first reason about why the extracted features can successfully predict song reuse across the platforms. *Signaling theory* [8] is often applied in selection scenarios such as choosing a song to collaborate with its author, where *signals* – i.e., observable pieces of information such as the number of an author's followers and the likes received by a song – are assessed to infer a hidden quality of the signaler – i.e., in this case, the author's talent and the quality of the song. We argue that the extracted features are used by community members as *assessment signals* [23], that is, metadata that make the music platforms transparent to the quality of both the artifacts and the talent of the authors. Previous research on OSS development has confirmed that observable signals in online profiles of both individuals (e.g. Stack Overflow reputation score [4], GitHub followers [18]) and projects (e.g., GitHub badges for passing builds and high test coverage [27]) are used as proxies for expertise, commitment, and quality. Our work complements these findings in the scenario of collaborative songwriting platforms.

**Popular Songs**

In our original study on Songtree, we found general support for the hypothesis H1 that 'popular' songs signaling many reactions (i.e., the number of like, bookmarks, and comments) have better chances to be derived. Overall, the findings from this study are in line with both the original study and prior research, thus reinforcing their validity. The effect of these features appears to be stronger in Splice, while ccMixter is similar to Songtree.

This finding complements those from prior research on other creative arts communities, such as Newgrounds [16, 17] and FAWM [2, 22], which did not evaluate the popularity of the artifacts as a predictor of future successful collaborations through reuse. Regarding OSS communities, the concept of 'popularity' does not apply to code contributions of collaborations between developers. Accordingly, Tsay et al. [28] studied the antecedents of change acceptance in GitHub, discussing the effect of popularity at the project level and using the number of stars (i.e., likes) and collaborators as proxies. Similarly, Jiang et al. [15] studied how repositories are forked in GitHub (i.e., copied from a developer's personal space into another's) and found that developers fork more often those owned by 'attractive' users.

The analysis of authors' responses to the questionnaire (see Table 6) show that the main driver for overdubbing a song is personal taste. By combining this result with the finding from the quantitative analysis of reactions (H1) in Table 5, we infer that authors signal appreciation by bookmarking rather than liking songs that they enjoy and want to reuse (`#bookmarks` OR=2.12, `#likes` OR= 1.18). Furthermore, a few authors reported that being mentioned (i.e., @username) in the comments to a song positively influences their decision to derive it. Future work should empirically validate the perceived positive effects of mentions on the probability of song reuse.

Hence, we confirm that popular artifacts are more likely to be reused both in arts and OSS communities.

**Recent and Derived Songs**

Regarding H2, we previously found very strong evidence (OR=0.02) that the longer since the upload of a song to Songtree, the less its chance to be overdubbed. Our findings on Splice (OR=1.43) and ccMixter (OR=1.34) are in contrast, indicating a positive association between the length of time a song is on a platform and the chances of being reused (i.e., a cumulative advantage for older songs). Consistently, the mean time to derive songs in Songtree is much smaller compared to Splice and ccMixter (see Table 4). The qualitative analysis offers an insight that might explain this difference. A few respondents report that the feed of updates showing friends' activity is one of the triggers of overdubbing in Songtree. Hence, we speculate that recent songs are strongly preferred in Songtree because they appear first in the update feeds, which are displayed in reverse chronological order and thus have a higher chance of capturing other members' interest, compared to older ones. On the contrary, the absence of a social network-like update feed in ccMixter may explain the different result. Regarding Splice, which does offer an update feed, its effect may have been limited by the lower mass of participants compared to Songtree.

Given these results, authors who want to foster the reuse of their own songs and increase their status in Songtree may benefit from adopting a strategy of maximizing the visibility of their songs by releasing them regularly, rather than in burst.

Regarding H3, in Songtree we found strong evidence (`song_depth` OR=0.56) that the more a song is distant from the root of the tree (i.e., builds on a chain of reused songs), the less likely it will be further overdubbed. Since Splice and ccMixter do not use the same song tree metaphor, we 'flattened' the feature to measure whether a song is the result of a derivation or rather a new one. Despite the change in the operationalization, the findings across the three communities concur to show an inverse relationship between the degree of derivativity and generativity (`is_derived` OR=0.47 and 0.11, respectively). Similar results on the loss of generativity in derived songs are reported by Cheliotis et al. [6], whereas Hill and Monroy-Hernandez [13] found that reused interactive media in the Scratch community are more generative than *de novo* content. In addition, a few survey respondents report that their decisions to reuse songs in Songtree depend on whether they see an opportunity to contribute. Hence, the combined evidence from these studies suggests that: (i) the inverse or direct relationship between generativity and derivativity depend on the type of the artifacts (e.g., songs vs. animations); (ii) in line with our initial expectation, for music artifacts the higher generativity of new songs depends on the fact that they leave more room for extension. The contrasting findings about the generativity of reused songs and animations are arguably explained by the different community goals. One of the reasons behind the success of Scratch its primary goal is introducing its members to software programming by giving them the means to learn. Instead, Songtree welcomes musicians and music lovers alike, but it does not provide any support to the latter in learning how to play and start contributing as songwriters. Thus, popular remixes in Scratch may be actually signaling to the community members that they should explore and reuse those aminations too if they want to learn the same things as many already did before. However, we point out that this is a mere speculation, which needs to be empirically validated in future work. Finally, this finding on the loss of song generativity is actionable in the sense that community members seeking to increase their status should either start new songs or reuse 'less derived' ones, thus leaving others enough space to build upon their own work.

**Author Status and Avatar**

In the original study, we found general support for our hypothesis H4 that songs by authors displaying high status in the Songtree community are more likely to be overdubbed. In this study, we find a consistent result with respect to the ccMixter community. Instead, the result regarding Splice is not statistically significant, arguably because of the limited size of its community. Unlike Songtree and ccMixter, with only a few hundred authors and spliced songs (see Table 1), Splice seems to have not reached a critical mass yet. Hence, one possible explanation for this contrasting finding may lie in the different dimensions of the communities.

In general, our findings on Songtree and ccMixter confirm the evidence from prior research on OSS communities, which found that the identity of developers is a strong antecedent of the likelihood of accepting their code contributions [12]. Still, it is interesting to note that, despite the strong positive effect of authors status revealed by the regression analysis, none of the authors who answered the survey (n=49) mentions author ranking as a trigger for overdubbing. On the contrary, two respondents call for entirely abolishing the ranking of authors from Songtree, which may allegedly boost rivalry rather than collaboration.

In Songtree and ccMixter, we find consistent support for the last hypothesis H5, according to which songs by authors easily recognized by their profile picture are associated with higher odds to be derived (H5 was not tested for Splice). Our finding contrasts with the work of Stanko [26], who found that promoting artists and 3D object designs on the front page of the Thingiverse community website has no impact on the likelihood of remixing. The diverse nature of the creative artifacts may again explain the contrasting findings.

Furthermore, our qualitative analysis reveals that the feed of activity of contacts is perceived by authors as one of the triggers that influence their choice to overdub a song. This result reinforces the evidence of the positive effect of having an avatar on the likelihood of song reuse. In fact, depending on the image, customizing the avatar may make it easier to identify authors and consequently reduce the chance of their last activity to go unnoticed in the feed of updates.

Overall, given the results of hypotheses H4 and H5, this study provides robust evidence that the status and the identity of community members are consistently used as proxies of artifact quality regardless of the domain, whether technical (as in OSS communities) or artistic (as in music communities).

Furthermore, these findings have actionable implications as authors can improve their status in these communities by making their profile more appealing and easier to recognize.

## 5.2 Implications for Researchers

The $R^2$ and AUC values in Table 5 indicate Songtree as the best model. Yet, the four models have been fit to different datasets extracted from different communities, so there is no measure (e.g., AIC) that allows us to make a direct comparison of goodness-of-fit. Still, when comparing the SongtreeRC model to Songtree, which share the same dependent variable, we observe a 35% drop in $R^2$ (0.31 vs. 0.89, respectively) and, in most cases, a reduction in the effect size for the significant predictors. These differences, which are arguably explainable through the only difference between the two datasets (i.e., feature extraction time), suggest that predictors extracted just before the event of interest are capable to explain more accurately the phenomenon of song reuse. Since the features in the datasets for ccMixter and Splice are not extracted at data collection time, we argue that the performance of these two models would increase if we could extract the predictors when a song is reused.

Unlike prior studies on OSS communities [9, 28] and with some notable exceptions such as the longitudinal study on Scratch by Hill and Monroy-Hernández [14], previous research on reuse in creative arts communities either overlooked the importance of extracting features at the instant of time before the event of interest or missed to report it. Hence, one lesson learned from this study is that future research on artifact reuse in creative arts communities should be more careful about how predictors are extracted when dealing with cross-sectional data.

### 5.3 Suggestions for Practitioners

The topics that have emerged from the analysis of the user survey allow us to provide some suggestions and insights potentially useful to the designers of both OSS and collaborative music platforms.

The analysis of the survey has revealed the importance of the social networking features already implemented in collaborative songwriting platforms. The friend networks are in fact leveraged by Songtree users for a variety of purposes, from song discovery to requesting extensions via overdubs. However, none of these platforms supports collaboration to an extent reached by OSS collaborative platforms. Much of the popularity of the social coding site GitHub is in fact due to the support of the 'fork-and-pull' collaborative development model,[7] which allows developers to copy (fork) projects to their personal space, freely apply changes with no coordination needed, and finally ask permission (pull request) to the owner of the original project for returning the changes. Visual features to preview and review proposed changes before accepting them are also available. Currently, none of the platforms studied provides comparable mechanisms to support and coordinate remote work for musicians. Only Splice provides basic implementations of song projects and membership. As such, compared to OSS development, the music co-creation process in these music platforms is more *ad hoc* than managed. Based on the qualitative analysis of the survey, we argue that the implementation of GitHub-like features for the management of virtual bands (i.e., similar to GitHub organizations) or virtual albums (i.e., similar to GitHub projects from the same organization), along with support for the versioning of audio material, has the potential to ease collaborative songwriting and, in turn, possibly increase the popularity of these communities.

Another topic that emerged from the survey is the inefficiency of the current genre taxonomy to browse the music catalog and discover songs for potential collaborations. According to survey respondents, the current approach of imposing a top-down taxonomy is not flexible enough. As such, community administrators should instead consider the use of a bottom-up approach, such as a folksonomy [24], letting users define genres and categorize songs through tags. ccMixter is already adopting free tags to label songs, however, they are used for displaying any kind of information about songs (e.g., audio format). A practical example of using a folksonomy to categorize artifacts is provided by Stack Overflow, which requires askers to label their questions using tags. However, these tags are suggested to foster convergence (i.e., to prevent the use of similar tags like `C#` and `C-sharp`), and new tags can be only created by experienced members who have a reputation score high enough to unlock the related privilege.[8] Music platforms should consider accommodating the needs of their users by adopting a similar approach to define a folksonomy of music genres, enforcing tag convergence and letting only experienced members with a high author rank define new ones.

Finally, the qualitative analysis of the survey also suggests some insights potentially useful to OSS platform designers, who could benefit from implementing some features borrowed from collaborative songwriting platforms. For examples, projects hosted at GitHub could benefit from using gamification features like platform badges (as offered in Songtree), which would signal the level of embeddedness and fecundity of the team members (i.e., developers' ranking). For example, Foucault et al. [10]. have already proposed to leverage similar gamification features to reduce and prevent technical debt in software projects. Also, similar to the contests available in these music communities, OSS platforms could allow projects to offer prizes (e.g., unlocked project privileges, free private repositories) to attract newcomers and increase their sustainability. However, both software and artistic platform designers should be aware that some of the survey respondents reported that they feel intimidated by contests as well as by the competition for increasing member status within the community. As such, contests and activities within these platforms should be arranged in a way that they also reward collaboration rather than just competition among their users. For example, software project maintainers could reward members for successfully mentoring newcomers in applying their first commit to the code base.

[7] https://help.github.com/articles/about-collaborative-development-models

[8] https://stackoverflow.com/help/privileges/create-tags

### 5.4 Research Roadmap

While our study furthers the understanding of which factors from prior work in both OSS and art communities are predictive of successful collaboration through artifact reuse, there are some limitations to it that leave several research questions open, which we intend to answer in future work.

#### Song quality

Our quantitative analysis includes several predictors of reuse defined at song level (e.g., the number of likes and bookmarks received). However, we have neglected to look into the effects of the 'quality' of a song on the probability of its reuse, that is, *are 'better' songs more likely to be reused?* While it is reasonable to expect a positive effect, the challenge is rather how to objectively capture and quantify song quality. In OSS research, the quality of source code can be assessed to some extent through the use of static analysis tools, which extract metrics and compare them against a set of rules and baselines to identify low-quality code with potential issues (e.g., methods longer than X lines are harder to comprehend and maintain). Consistently, the qualitative analysis performed on Songtree in this study revealed that authors consider *sound quality* a prerequisite to overdub a song. Thus, in future work, we shall consider including in our model some predictors that capture, for example, the *encoding quality*, such as the bit rate in kbps and the use of lossy rather than lossless formats (e.g., MP3 and FLAC, respectively).

#### Collaboration patterns and needs

In the previous subsection, we have speculated on the results of the qualitative analysis and provided in particular suggestions to the designers of collaborative songwriting platforms for implementing GitHub-like features that would enable the organization of virtual bands and the management of song projects in a more controlled and structured way, similar to OSS teams and projects. We acknowledge that this is so far just a suggestion that still needs to be empirically validated in future work by answering the following research question: *does collaborative songwriting needs the same level of control and organization of collaborative software development?*

OSS projects typically adopt a clear ownership and management model by distinguishing owners from contributors – e.g., owners can reject contributions that are not useful, correct, or of sufficient quality.

Among the music communities studied in this paper, Splice is already adopting a collaboration model closer to this one. It already supports project management for songs, with members explicitly invited by virtual band leaders. Repositories can be freely spliced (i.e., forked) by anyone, but there is not yet a pull-request-like mechanism available to modify the parent song by merging back an external contribution.

Instead, in Songtree and ccMixter the current pattern of collaboration is less structured and controlled – everyone can reuse public songs by uploading extensions, regardless of their quality. Still, in our interactions with the project manager of Songtree, he has mentioned in multiple occasions that prolific and highly reputed authors have repeatedly requested the implementation of some features that would allow them to remove poor quality overdubs – as if the low quality of derived songs somehow diminished their own work.

Hence, there seems to be a potential for applying OSS-like solutions to manage music projects, but the circumstances under which similar patterns of team organization and control may be needed in the collaborative songwriting domain are still unclear. Accordingly, as future work, we intend to focus on studying communities such as Scratch and OpenProcessing,[9] which can be seen as hybrids of OSS coding and artistic collaboration/sharing. Moreover, these communities also have different notions of success and models of collaboration, which can also help us understand how current results generalize to other creative communities.

## 6 CONCLUSIONS

In this paper, we applied mixed methods to investigate co-creation behavior in three online music communities and determine which success factors from prior research in both OSS and online creative communities are antecedents of reuse.

Our findings show that, in line with previous findings from research on both OSS and creative communities, the perceived quality of an artifact in terms of users' reactions and authors' prominence heavily influences the choice to reuse an artifact. We also observed a loss of generativity and interest from other artists in songs that have been already reused. As for the authors' popularity, we found that social status is used to signal to fellow artists the quality of artifacts, which in turn see an increase in generativity.

Finally, we provided some suggestions about features that arts and OSS online communities may consider borrowing from each other.

## ACKNOWLEDGMENTS

We are grateful to Songtree, for opening their data for research purposes, and to the users who participated in the survey.

[9] www.openprocessing.org